\documentclass[10pt,journal]{IEEEtran}
\usepackage[nocompress]{cite}
\usepackage[mathscr]{euscript}

\usepackage{stfloats}
\usepackage{graphicx}
\usepackage[cmex10]{amsmath}
\usepackage[ruled]{algorithm2e}
\usepackage{algorithmic}
\usepackage{array}
\newcolumntype{P}[1]{>{\centering\arraybackslash}p{#1}}
\newcolumntype{M}[1]{>{\centering\arraybackslash}m{#1}}
\usepackage{mdwmath}
\usepackage{mdwtab}
\usepackage{url}
\usepackage[english]{babel}
\usepackage{amsfonts}
\usepackage{mathbbol}
\usepackage{amssymb}
\usepackage{amsthm}
\usepackage{mathtools}
\usepackage{adjustbox}
\usepackage{comment}
\usepackage{multicol}
\usepackage{multirow}
\usepackage{placeins}
\usepackage{dsfont}
\usepackage{soul,color}
\usepackage{tabu}
\usepackage{multirow}
\usepackage{booktabs}
\usepackage{cleveref}
\usepackage{cite}
\usepackage{paracol}
\usepackage[many]{tcolorbox}
\usepackage{xcolor}
\usepackage{physics}
\usepackage{enumitem}
\usepackage{wrapfig}

\NewTColorBox{NewBox}{ s O{!htbp}  }{%
	floatplacement={#2},
	IfBooleanTF={#1}{float*,width=\textwidth}{float},
	colframe=black!50!black,colback=white!95!black,valign=center, halign=flush center,enhanced,attach boxed title to top center={yshift=-3mm,yshifttext=-1mm},colbacktitle=black!70!white,
	title=Objectives,fonttitle=\bfseries,
	boxed title style={size=small,colframe=black!50!black}
}

\usepackage{float}
\usepackage{etoolbox}
\makeatletter
\patchcmd{\@makecaption}
{\\}
{:\ }
{}
{}
\patchcmd{\@makecaption}
{\scshape}
{}
{}
{}
\makeatother

\DeclarePairedDelimiter\floor{\lfloor}{\rfloor}

\SetAlgoCaptionLayout{centerline}

\begin{document}
	
	\title{SPoTKD: A Protocol for Symmetric Key Distribution over Public Channels Using Self-Powered Timekeeping Devices}
	
	\author{Mustafizur~Rahman,~ Liang Zhou,~
		and~Shantanu~Chakrabartty,~\IEEEmembership{Senior~Member,~IEEE}
		
		\thanks{M. Rahman and S. Chakrabartty are with the Department of Electrical and Systems Engineering, Washington University in St. Louis, St. Louis, Missouri 63130, USA and L. Zhou is with Analog Devices Inc. All correspondences regarding this manuscript should be addressed to shantanu@wustl.edu.}
		\thanks{This work is supported in part by a research grant from the National Science Foundation CNS-1646380}}
	\maketitle
	
	\begin{abstract}
		
		In this paper, we propose a novel class of symmetric key distribution protocols that leverages basic security primitives offered by low-cost, hardware chipsets containing millions of synchronized self-powered timers. The keys are derived from the temporal dynamics of a physical, micro-scale time-keeping device which makes the keys immune to any potential side-channel attacks, malicious tampering, or snooping. Using the behavioral model of the self-powered timers, we first show that the derived key-strings can pass the randomness test as defined by the National Institute of Standards and Technology (NIST) suite. The key-strings are then used in two SPoTKD (Self-Powered Timer Key Distribution) protocols that exploit the timer's dynamics as one-way functions: (a) protocol 1 facilitates secure communications between a user and a remote Server; and (b) protocol 2 facilitates secure communications between two users. In this paper, we investigate the security of these protocols under standard model and against different adversarial attacks.  Using Monte-Carlo simulations, we also investigate the robustness of these protocols in the presence of real-world operating conditions and propose error-correcting SPoTKD protocols to mitigate these noise-related artifacts. 
		
		
	\end{abstract}
	
	\section{Introduction}
	\label{intro}
	\IEEEPARstart{S}{ecuring} information exchange with internet-of-things (IoTs), is becoming ever more important due to the proliferation of these platforms in domains ranging from infrastructure-IoTs ~\cite{aonoiscas} to medical-IoTs~\cite{mediot}. In one study~\cite{iot42} it is claimed that around 98\% of the IoT data traffic is unencrypted and hence vulnerable to a data breach. Conventional data encryption techniques like RSA are too computationally prohibitive to be universally implemented on these low-resource platforms and reducing the computational complexity makes the approach vulnerable to quantum attacks. For instance, it is estimated in literature that a quantum computer with 8194 logical qubits using Shor’s Algorithm would be able to break the Rivest-Shamir-Adleman(RSA)\cite{Riv78} system with a key size of 4096 bits in 229 hours while for Discrete log problem with a key size of 521 bits it would take 55 hours for a quantum computer with 4719 logical qubits, again using the Shor’s Algorithm\cite{fourquant}. Symmetric key algorithms like Advanced Encryption Standard (AES-256) can be customized for IoT platforms and are considered to be secure against quantum attack~\cite{fourquant}, provided the security of the initial key-exchange can be guaranteed. Quantum key distribution(QKD)\cite{Ben84} which is based on the principles of quantum-mechanics, like quantum entanglement \cite{Eke91} or the no-cloning principle \cite{Gis02},\cite{por14} could be used to guarantee the security of the initial key-exchange. However, one of the major drawbacks of current state-of-the-art QKD systems is that they require dedicated and specialized peer-to-peer communication links \cite{Ngu06},\cite{Ngu08},\cite{Kor14},\cite{Yin17}. Not only do these links require careful maintenance and calibration to ensure quantum-coherence, but these systems are also expensive and not portable. Hence, current QKD systems cannot be scaled for internet-scale key distribution \cite{Jai11},\cite{Bra00} and communications involving lightweight IoT devices with resource constraints will still be vulnerable to quantum attacks.
	\begin{figure}
		\begin{center}
			\includegraphics[page=1,scale=0.39]{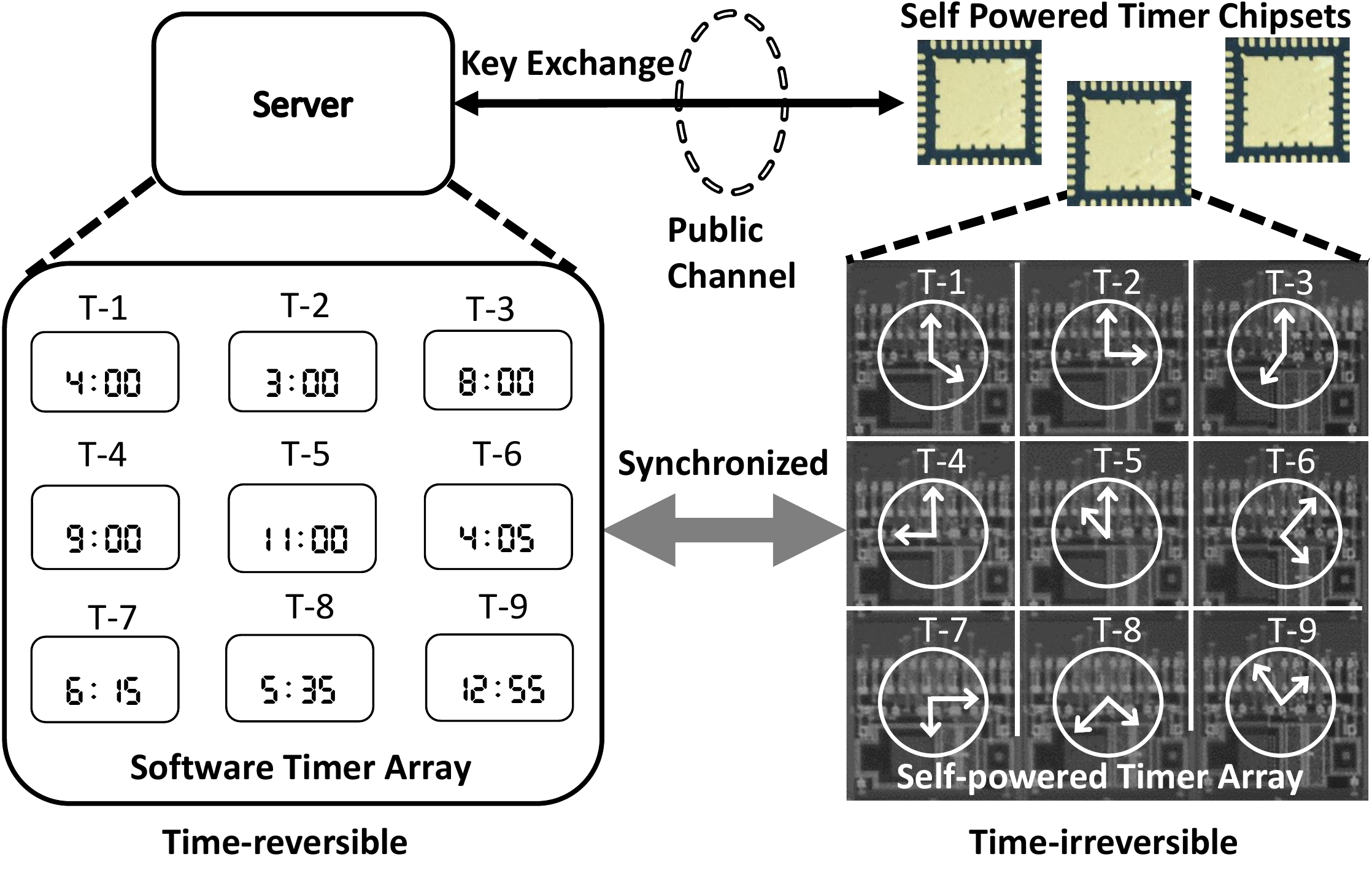}
		\end{center}
		\caption{Framework underlying SPoTKD protocols: the synchronization and time-irreversibility of self-powered timers is exploited to implement one-way functions and facilitate secure key exchange over public channels.}
		\label{fig_motivation}
	\end{figure}
	\par In this paper, we propose a hardware-software Self-Powered Timer based Key distribution (SPoTKD) framework that does not require any modifications to the existing communication infrastructure, can be scaled to a large number of IoTs, and is potentially secure against quantum attacks. The approach relies on the trend that silicon-based chipsets with the capability of integrating billions of transistors and memory elements~\cite{million} can be manufactured in a large scale and at a low-cost~\cite{cost}. If a physical feature on these chipsets could be exploited to implement a secure one-way function, then a hardware-software approach could be used to support key distribution over public channels. In this paper, we propose one such method that exploits the synchronization capabilities and security features of our previously reported \cite{Zho17a} self-powered timekeeping devices. The basic framework for SPoTKD is illustrated in Figure~\ref{fig_motivation} where multiple identical copies of self-powered timer chipsets are openly distributed to all the users. Each of the timers on these chipsets is synchronized with its software clone running on a server. The key exchange between the server and the user is achieved based on this synchronization and time-evolution is used to implement a secure one-way function. It is to be noted that once the secret keys have been established and exchanged between the two parties, traditional symmetric cryptographic algorithms can be used for secure communications and user authentication\cite{Ngu07}.
	\par The rest of this paper is organized as follows. Section \ref{relate} briefly describes other related protocols based on hardware-software based key distribution. Section \ref{back} provides a brief background of the previously reported self-powered timers and their essential security features that have been exploited in the design of the SPoTKD protocols. In Section \ref{keygen}, we propose two SPoTKD protocols, one between a server and any user, and the other between two users. In Section \ref{security} we analyze the security of the proposed protocols under various adversarial attacks. The robustness of protocol to operating and hardware artifacts have been analyzed in Section \ref{noise} and in Section \ref{errcor} we introduce a variant of the protocol that uses error-correction codes to improve noise-robustness. We conclude the paper in Section \ref{disc} with discussions about the challenges and future directions.
	
	\section{Related Works}
	\label{relate}
	In literature, a few hardware-software key exchange methods have been proposed. In \cite{rabin_key} a hardware-software public-key cryptography system for wireless networks was proposed based on Rabin's Scheme \cite{rabin}. However, the security of Rabin's Scheme relies on the difficulty of factorizing large numbers, hence, it has similar vulnerabilities as the classical DH or RSA methods. Meanwhile, the one-way function (time irreversibility) implemented in SPoTKD is based on the principle of physics. Thereby SPoTKD does not suffer from such vulnerabilities. In \cite{perfect} a hardware-software key exchange technique was proposed that exploited correlations across chaotic wavepackets in classic optical communications channels. However, the method still requires peer-to-peer connectivity between the users and hence has similar scaling disadvantages as QKD methods. On the other hand, SPoTKD uses silicon-based chipsets containing self-powered timers. Therefore, SPoTKD has the advantage against such key distribution methods for platforms with low computational resources. The hardware-software approach proposed in \cite{chaos} used chaos synchronization to distribute random keys over public channels. However, due to the lack of reliable synchronization, this approach incurs significant errors during decryption. Recently, Physical Unclonable Function(PUF) based hardware-based encryption key distribution has been proposed. A specific variant of this technique, described in~\cite{PPUF} as Public Physical Unclonable Function(PPUF) has been used for public-key cryptography and leverages the difficulty of accessing physical information stored on chipsets. However, in PPUF the stored information is static in nature and hence is potentially vulnerable to machine learning attacks \cite{SVM,DL}. Whereas in SPoTKD the keys are derived from dynamic information that changes with time.
	\begin{figure*}
		\begin{center}
			\includegraphics[page=1,scale=0.68]{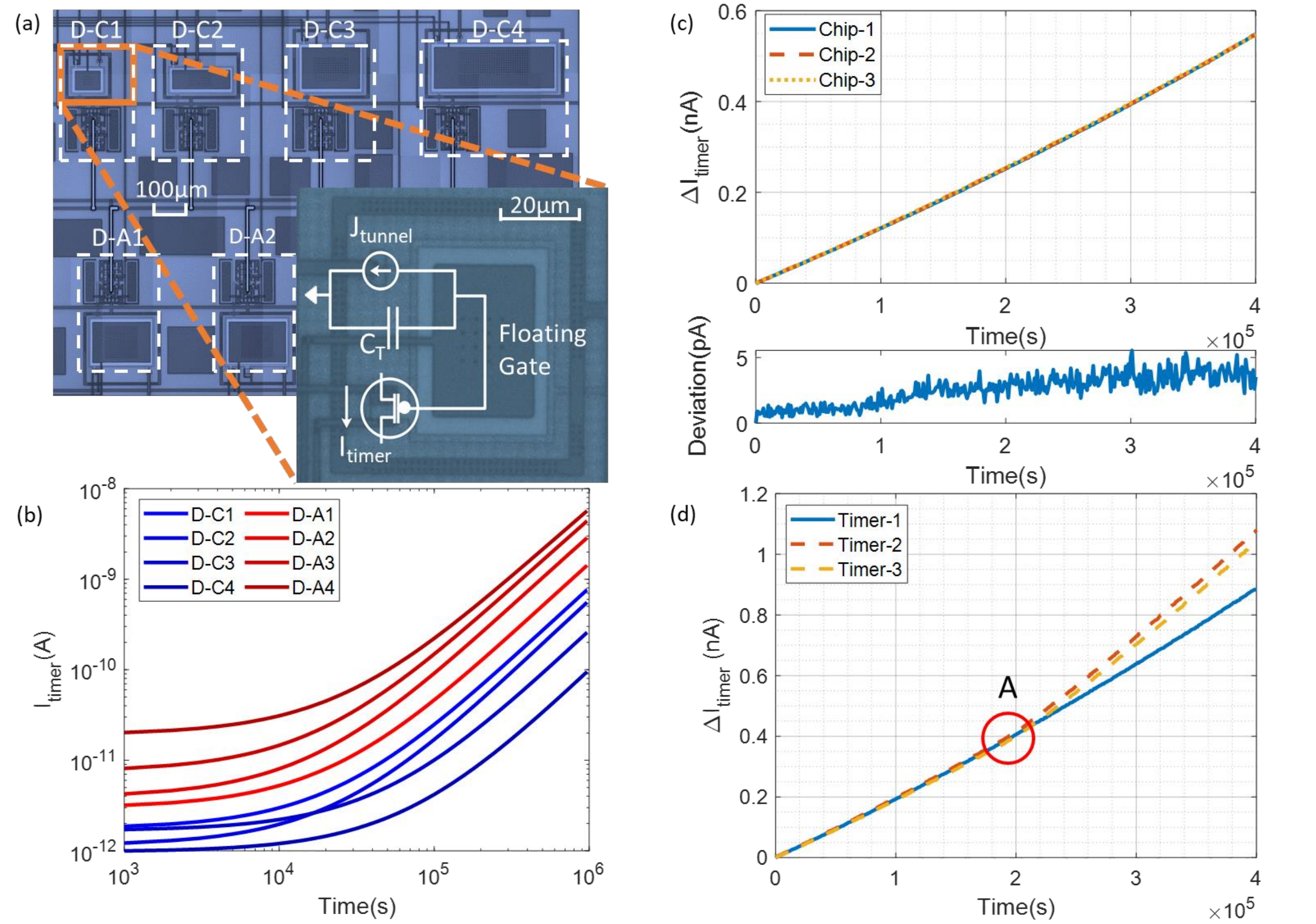}
		\end{center}
		\caption{ (a) Micrographs of self-powered timers (labeled as D-C1, D-C2, D-C3, D-C4) with different form factors and features that determine the parameters of the timer behavioral model in equation~\ref{eqn:readout}. The equivalent circuit model for a single timer along with the readout circuit is shown in the inset. (b) The temporal responses measured using these timers for (a) different initialization conditions. (c) Synchronization of a timer's temporal response with same form factors across multiple chipsets after the initial transient response. (d) Desynchronizing the temporal response of different timers by coupling an external source of energy into two of the timers at the time-instant denoted by A.}
		\label{fig_prel}
	\end{figure*}
	
	\begin{figure}
		\begin{center}
			\includegraphics[page=1,scale=0.63]{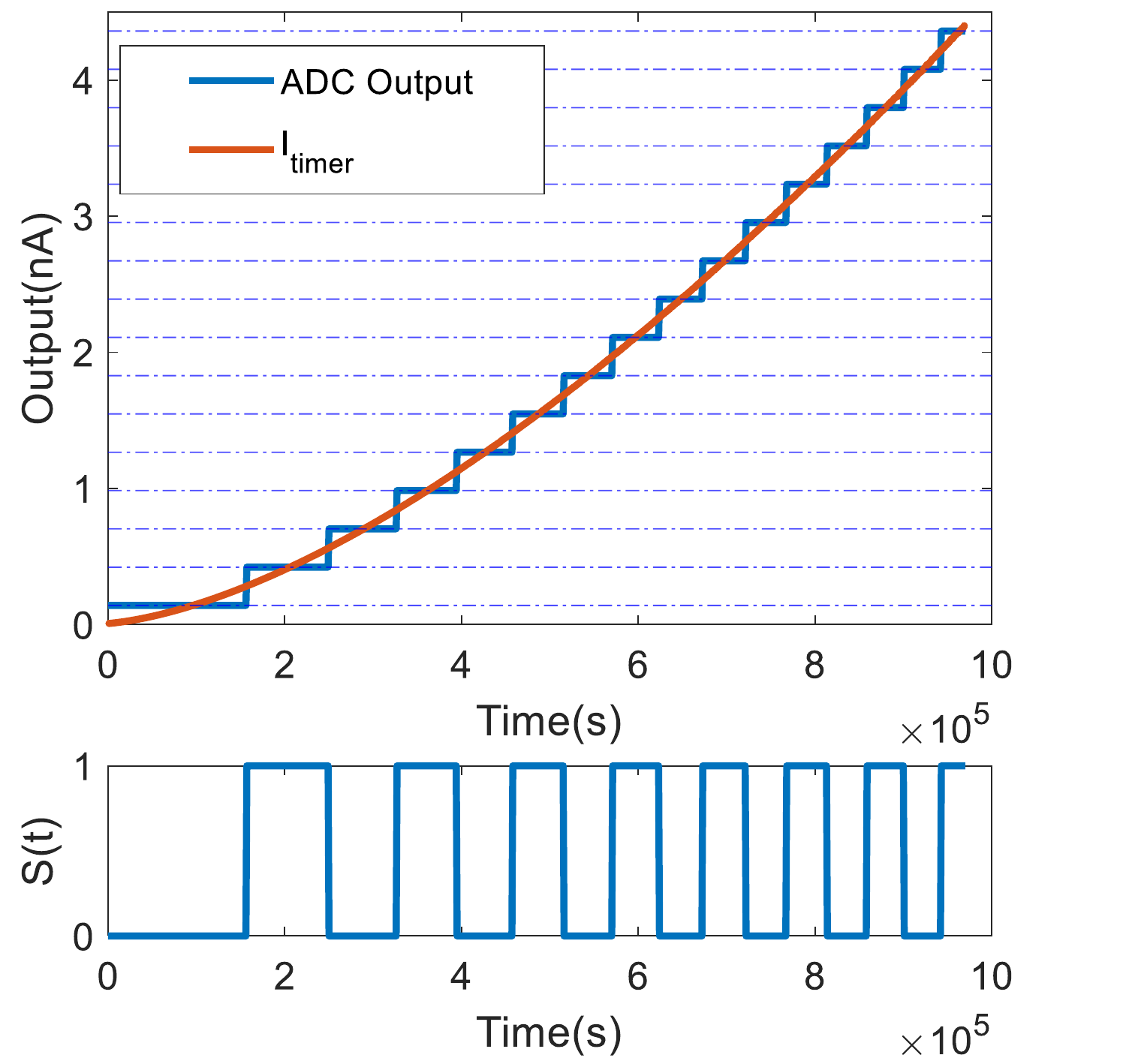}
		\end{center}
		\caption{Dynamic binary state $s(t)$ of a timer generated after the analog current is read-out with an ADC. Illustration here shows the state $s(t)$ corresponding to a 4-bit ADC.}
		\label{fig_quant}
	\end{figure}
	
	\begin{figure*}
		\begin{center}
			\includegraphics[page=1,scale=0.52]{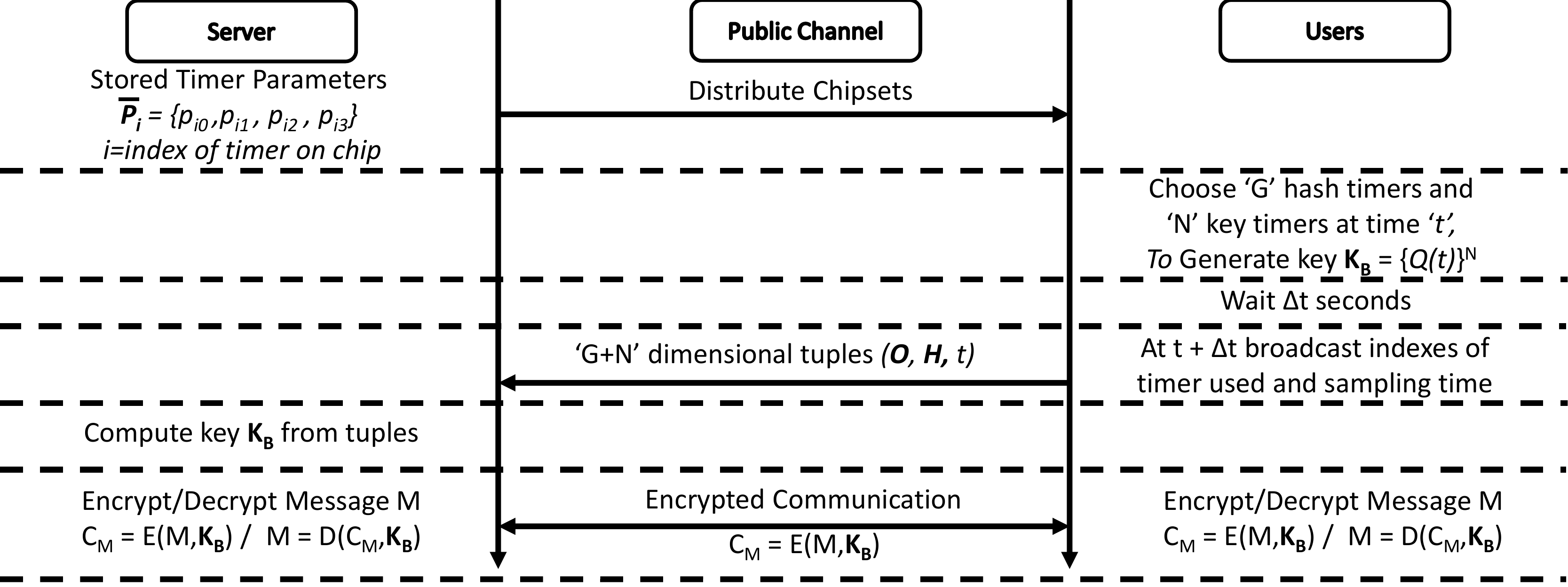}
		\end{center}
		\caption{Basic SPoTKD protocol between the server and a user. Here $E(M,\mathbf{K_B}$) represents an encryption function where message $M$ is encrypted with key $\mathbf{K_B}$, $D(C_M,\mathbf{K_B})$ is the decrypting function where $C_M$ is the cipher text being deciphered with key $\mathbf{K_B}$.}
		\label{protocol_1}
	\end{figure*}
	
	\section{Self-powered Timer Security Primitives}
	\label{back}
	The SPoTKD protocol exploits the physical features of self-powered timers to ensure the security of the key exchange. The design and the operating principle of self-powered timers have been previously reported in \cite{Zho17a,Zho19}. In this section, we discuss the basic security primitives offered by the timer's physical response that will form the axiomatic core of the security analysis for SPoTKD that is presented later in this paper.
	\subsection{Self-powered timers are immune to power side-channel attacks}
	A simplified equivalent circuit model of the self-powered timer is shown in Fig~\ref{fig_prel}(a) where a leakage-current $J_{tunnel}$ is used to discharge a floating-gate capacitor $C_T$. Thus, once the floating-gate capacitor $C_T$ is charged or programmed initially, no external power is required to drive the dynamics of the discharge process. The change in the floating-gate charge/voltage is monotonic with respect to the time elapsed and this feature has been previously used for time-keeping, synchronization, and authentication~\cite{Zho17a,afifi}. For this work, the self-powered operation decouples the timer from the external power supply. This provides security against any power side-channel attack that might be aimed at gaining knowledge about the current state of the timer by observing fluctuations in the supply-current.
	\subsection{Self-powered timers are immune to electromagnetic side-channel attacks}
	The leakage current $J_{tunnel}$ in the self-powered timer is implemented using Fowler-Nordheim(FN) tunneling of electrons through a thin gate-oxide barrier. In~\cite{Zho19}, we have shown that the operation of the timers is robust even when the FN tunneling current is as low as one electron per second (or less than an attoampere). From a security point of view, the low
	tunneling current practically eliminates any electromagnetic (EM) emission and hence any EM side-channels. Also, any unauthorized attempt to access the timer-state using an EM probe desynchronizes or destroys the state of the timer.
	
	\subsection{Dynamics of the self-powered timers can be synchronized}
	One of the essential attributes of the timer that is important for the realization of the SPoTKD protocol is that the timer's temporal responses can be synchronized not only with respect to each other but also to a well-defined behavioral (or software) model. For this work, we use a specific form of the timer behavioral model that is given by
	\begin{equation}
	I_{timer}\small(t\small)=p_3\exp\Big[-\frac{p_2}{\log \small(p_1t + p_0\small)}\Big].
	\label{eqn:readout}
	\end{equation}
	where $I_{timer}(t)$ is the current measured at time instant $t$ quantifying the state of the timer. The current is measured using a read-out metal-oxide-semiconductor field-effect transistor (MOSFET) whose gate is coupled to the floating-gate, as shown in Fig.~\ref{fig_prel}(a). The behavioral model in equation~\ref{eqn:readout} assumes that the read-out transistor is biased in a specific regime, details of which can be found in the derivation of the behavioral model in the Appendix. The tuple $\overline{P} = [p_0, p_1, p_2, p_3]$ in equation~(\ref{eqn:readout}) are the timer parameters that are determined by the device form factors and the device initialization conditions. Figure~\ref{fig_prel}(a) shows an example of a system-on-chip implementation that integrates different timer structures with varying form-factors. The responses of these timers with different initialization conditions are presented in Figure~\ref{fig_prel}(b) which shows that the temporal dynamics of each timer is unique and is determined by the tuple $\overline{P}$. We have previously shown that for a fixed set of timer parameters $\overline{P}$ the mathematical model in equation~\ref{eqn:readout} can capture the temporal behavior of the timer for more than a year with an accuracy of greater than 0.5\% \cite{Zho19}. This is shown in Figure~\ref{fig_prel}(c), where the timers with the same form-factor but integrated on different chipsets remain synchronized with each other. The deviation between the timer's responses is in the range of pico-amperes and this synchronization error can be attributed to the measurement noise and not to the synchronization error. For the SPoTKD protocol, the synchronization between the behavioral model (or software timer) and the hardware timers will be used for key exchange. The key exchange will exploit the asymmetry between the software timers and hardware timers where that the hardware timer cannot be rewound (or time-irreversible) whereas its software clone can be rewound to any previous time instant. This asymmetry is exploited as a one-way function for securing the SPoTKD protocol. Note that the parameters $\overline{P}$ which determine the dynamics of each timer, are never revealed publicly and therefore functions as a private key in our protocol. Later in Section~\ref{security} we show that it is practically impossible to extract these parameters from measurements on the hardware timer itself.

	\subsection{Self-powered timers are designed for one-time read and tamper-resistant}
	\label{sec:onetimeread}
	 In~\cite{Zho19} we showed that the synchronization between the timers could be broken by injecting an external signal into the floating-gate. This is demonstrated in Figure~\ref{fig_prel}(d) where three timers (with similar form factors) are synchronized with respect to each other till time-instant 'A'. Then at time-instant 'A' an external energy-source is coupled to timers 2 and 3 (in this case using capacitive coupling). As a result, these timers become de-synchronized from each other. We will use this controlled de-synchronization feature to intentionally destroy the dynamical state information stored on each timer once its state has been accessed. Thus, each of the timers can only be used once to generate the key-string after which the state of the timer is destroyed (or desynchronized). Note, the desynchronization of the timer can also result when the timer is unintentionally probed (using hardware delamination or using electromagnetic probing). This feature makes the basic timer tamper-resistant.
	  
	\subsection{Bit generation using self-powered timer}
	\par We will assume that the state of the self-powered timer can be measured using an on-chip analog-to-digital converter(ADC) where the least-significant-bit (LSB) represents a modulo-2 measurement of the timer value. Denoting the binary state $s(t) \in \{0,1\}$ of the timer as the LSB obtained after the $I_{timer}(t)$ is measured at a time-instant $t$, then $s(t)$ can be expressed as
	\begin{equation}
	s\small(t\small)=\floor{\frac{I_{timer}\small(t\small)}{\delta}}\mod 2
	\label{eqn:bitgen}
	\end{equation}
	where $\delta$ is the resolution of the ADC. This is illustrated in Figure~\ref{fig_quant} where a 4-bit ADC is used to measure $I_{timer}(t)$ to generate the LSB or $s(t)$. For the protocols proposed in this paper, we will also assume that once the binary state of a timer is measured, its state is destroyed through a process of desynchronization, as described in the section~\ref{sec:onetimeread}. This implies that each timer can only be used once to generate a single bit '$s(t)$' at a given time $t$ for key-generation. 

	\subsection{Summary of hardware security primitives offered by self-powered timers}
	\label{sec:primitives}

	\par Here we summarize the security primitives that is offered by self-powered timers and will serve as axioms for the proposed SPoTKD protocol: 
	\begin{enumerate}
		\item[SP1:] It is practically impossible to access any information about the secret parameters or the state of the timer using side-channels (power or electromagnetic) attacks.
		\item[SP2:] The temporal behavior of each timer is unique and are determined by the timer's secret parameter tuple $\overline{P}$.
		\item[SP3:] The binary state of a timer $s(t)$ as defined in equation~\ref{eqn:bitgen} is dynamic in nature and changes with time. As a result, the state of a timer is unpredictable without knowledge about the secret parameters of the timer.
		\item[SP4:] The hardware chipsets are designed in such a manner so as users are limited to only the output of the chipsets after following specific protocol (discussed in Section IV). Any attempt to snooping on the hardware chipsets otherwise would result in a destruction of the information embedded on the timers.
		\item[SP5:] The state of each timer in a chipset can only be accessed once, after which the state is erased or destroyed.
		\item[SP6:] The number of hardware chipsets that are available at any given instance of time is finite.
	\end{enumerate}
	
	\section{SPoTKD Protocol}
	\label{keygen}
	\begin{figure*}[b]
		\begin{center}
			\includegraphics[page=1,scale=0.5]{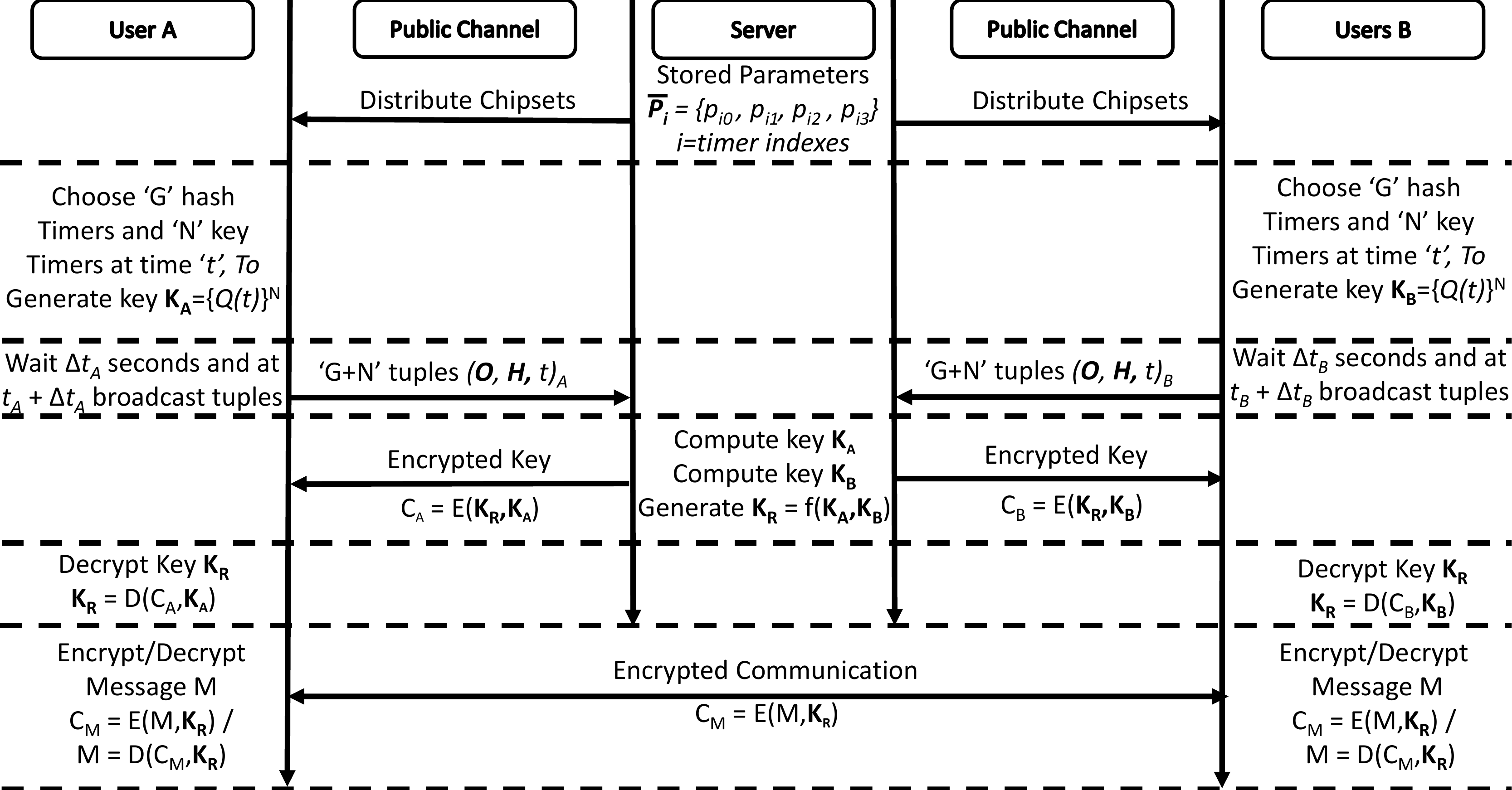}
		\end{center}
		\caption{SPoTKD protocol for exchanging keys between two users with server acting as a trusted third party. }
		\label{protocol_2}
	\end{figure*}

	
	
	\par The basic SPoTKD protocol is shown in Figure~\ref{protocol_1}. A server creates multiple replicas of chipsets each of which integrates a set $\mathcal{T}$ of $C \in \mathbb{Z}^+$ timers. Each timer in the set is assumed to be initialized according to a parameter tuple $\overline{P}_i$, where $ 1 \le i \le C$, as defined in equation~(\ref{eqn:readout}). Note that some of the parameters (initial charge on the floating-gate) in the tuple are programmed by the server and some of the parameters (device form-factor) are fixed post-fabrication. Also, note that only the server has access to this information and is kept secret from the users. These identically programmed chipsets are then distributed to all the users over a public distribution channel, as shown in Fig.~\ref{fig_motivation}. When an intended user wishes to communicate with the server, they arbitrarily choose to measure the binary states of two sets of timers which will be referred to as 'hash' timers and 'key' timers. The objective is to use the $G$ 'hash' timers and $N$ 'key' timers to generate an $N$ bit long binary key $\mathbf{K_B} \in \{0,1\}^N$. To achieve this 
	the outputs of $G$ randomly chosen hash timers $s_{H_1}(t),..,s_{H_G}(t)$, $ 1 \le H_1,..,H_G \le C$ measured at time instant $t$ are XOR-ed with each other to generate a single bit $X(t)$ according to 
	\begin{equation}
	\begin{array}{cc}
	X(t)=s_{H_1}(t)\oplus s_{H_2}(t) \oplus s_{H_3}(t).....\oplus s_{H_G}(t)
	\end{array}
	\label{eqn:keybitgen}
	\end{equation}
     Note that the time instant $t \in \mathbb{R}^+$ is referenced according to a universal standard time.
	 The key bits $Q_{L}(t), L = 1,..,N$ are then generated at time $t$ by XOR-ing the binary states of each of the 'key' timers $s_{O_1}(t),..,s_{O_N}(t); 1 \le O_1,..,O_N \le C$ with $X(t)$ according to 
	 \begin{equation}
	 \begin{array}{cc}
	 Q_{L}(t)=s_{O_L}(t)\oplus X(t)
	 \end{array}
	 \label{eqn:keystring}
	 \end{equation}
	to generate $\mathbf{K_B} = \{Q_L\}^N$.Note that since the state of each of the timers can only be accessed once, the `hash' and the 'key' timers need to be different, namely $\{O_1,..,O_N\} \cap \{H_1,..,H_G\} = \emptyset$. Also, note that the user can only access the $N$ bit key string $\{Q_L\}^N$ and not the binary states of the 'key' timers or $X(t)$ from the hardware chipsets.

	In the next step of the SPoTKD protocol, as shown in Figure~\ref{protocol_1}, the user waits for a random time-duration $\Delta$t seconds after which they broadcast a $G+N$ dimensional tuple $(\mathbf{O},\mathbf{H},t)$ over the public channel. Note that here $t$ indicates the time at which the $G$ 'hash' and $N$ 'key' timers were accessed and only the indices of the timers are broadcasted (and not measured output). The server then uses the tuples $(\mathbf{O},\mathbf{H},t)$ and its knowledge of the 'secret' parameters $\overline{P_i}, 1 \le i \le C$ to decipher the binary states of all these timers and compute the key $\mathbf{K_B}$ completing the key exchange.

	\par The SPoTKD protocol shown in Figure~\ref{protocol_1} is suitable for communicating between a user and a server that owns and initializes all the timer chipsets. However, key exchange between two users can also be facilitated with the help of the server acting as a trusted third party, as shown in Figure~\ref{protocol_2}. In this protocol, both the users broadcast their tuples $(\mathbf{O},\mathbf{H},t)_A$ and $(\mathbf{O},\mathbf{H},t)_B$ over a public channel. The server deciphers both keys, $\mathbf{K_A}$ and $\mathbf{K_B}$ according to previous protocol. The server then generates a new key $\mathbf{K_R}$ which is a function of the keys $\mathbf{K_A}$ and $\mathbf{K_B}$. This function $f:\{0,1\}^{2N} \rightarrow \{0,1\}^N$ is decided by the server and can be any mathematical operation ranging anything from multiplication to complex hashing. This operation is never revealed and changed for every session. The server then sends cipher texts $C_A = E\small(\mathbf{K_R}, \mathbf{K_A}\small)$ to user A and $C_B = E\small(\mathbf{K_R},\mathbf{K_B}\small)$ to user B containing the key $\mathbf{K_R}$ encrypted using $\mathbf{K_A}$ and $\mathbf{K_B}$ respectively. The users can decrypt the cipher text to know the secret key $\mathbf{K_R}$. For further communication, each user uses this key $\mathbf{K_R}$ to encrypt and decrypt their messages with each other. Since all keys are randomly generated and have never been used before then anyone intercepting the cipher text will not gain any information regarding the secret key being used. Note that in this protocol the users do not need to match either the timers they used in the chip or the time at which they will generate their respective keys. They only need to agree upon their time of communication and can generate their keys beforehand individually. In order to update any new session key between two users, the users would need to use a new set of timers and follow the same protocol for exchanging keys with the server acting as the trusted third party.
	
	\section{Security and Performance Analysis}
	\label{security}
	\par  According to the recommendation by National Institute of standards and technology (NIST), a 256-bit key is sufficient for symmetric key algorithms to be secure \cite{keylen} even in the presence of a quantum computer. While using Grover’s algorithm, a quantum computer with 6681 logical qubits and approximately $3.36\times 10^7$ physical qubits would require around $2.29\times10^{32}$ years for a brute force search attack on AES-GCM cryptosystem with a 256-bit key size \cite{fourquant}. Hence, for the rest of the paper, we will show test results corresponding to 256-bit keys, generated from $G = 128$ hash timers and $N = 256$ key timers for all analysis purposes. Note that, the number of hash timers used in key generation determines the complexity of the key generation. We will show that $G = 128$ hash timers are sufficient for the protocol to be secure. Increasing the number of hash timers would further increase the complexity but would come at the cost of noise robustness. Since the scope of this work is only to propose a secure key exchange protocol that can be used for symmetric-key encryption schemes, our security analysis will only focus on showing that the key exchange protocol is quantum secure.
	\begin{figure}
		\begin{center}
			\includegraphics[page=1,scale=0.45]{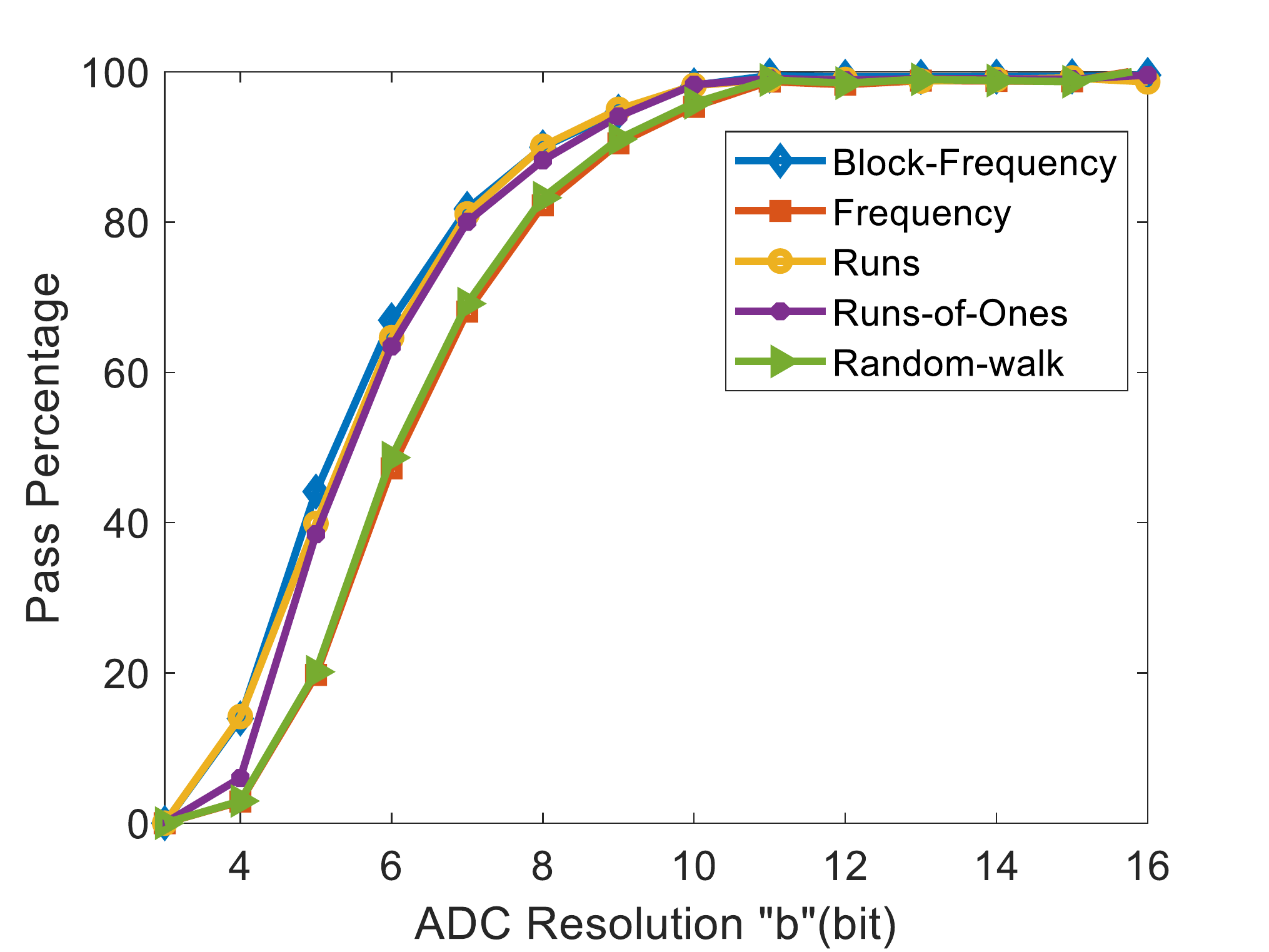}
		\end{center}
		\caption{Pass percentage obtained using the NIST randomness test suite applied to the keys generated using the SPoTKD protocol, as a function of the resolution $b$ of the ADC used to measure the state of the 'key' timers.}
		\label{fig_random}
	\end{figure}
	\par For our first analysis, we consider the scenario where an attacker simply attempts to guess the key without any information about the key generation system. As long as the keys that are generated from the timers are completely random in nature, the attacker will not gain any unfair advantage. So we tested the secret keys generated according to the SPoTKD protocol described in Section \ref{keygen} with the NIST test suite for checking randomness of bit stream\cite{nist}. The suite usually consists of fifteen different tests to measure the randomness in a certain bitstream. However, a few of these tests require a large sequence of bitstream which does not apply for a length of 256-bit keys. Therefore, in our analysis we show the test result for 5 of the suitable tests. The binary states for the hash timers were always measured with an 11-bit ADC irrespective of the key timers. This was performed to ensure better noise robustness. If a higher resolution ADC was used to sample the hash timers, then the noise robustness of the protocol would decrease (discussed in Section VI). Using Monte Carlo simulations, we sampled $10^6$ keys at random time instances using a b-bit ADC (or $2^b-1$ level quantizer) for the key timers. We extracted the parameter tuples $\overline{P} = [p_0, p_1, p_2, p_3]$ from the timer responses shown in Figure 2(b) and then randomized within the range of these actual hardware parameters to represent unique timers in our simulations. This ensures that each timer used in the simulation can actually be realized in hardware chipsets.
	Figure~\ref{fig_random} shows the pass percentage, i.e. the percentage of keys from the $10^6$ samples that passed the test, as the resolution 'b' of the ADC is varied for the key timers.
	We can observe from the plots that for large values of 'b', almost all the generated keys pass the test. The randomness degrades for ADC resolution less than 8 bits showing that a 9-bit ADC for the key timers should be sufficient to generate high-quality keys. This shows that keys derived from the timer responses are completely random in nature and any attempt to guess the key would result in a brute force search which is the same as breaking the AES-256 encryption scheme discussed above. Moreover, it also means that the binary state of each timer is uncorrelated with other timers and an attacker cannot simply sample the binary states of any one timer and can predict what other timers’ response would be at any given point in time. This is in accordance with the axioms SP2 and SP3 discussed in Section ~\ref{back}. 
	
	
		\begin{figure}
		\begin{center}
			\includegraphics[page=1,scale=0.45]{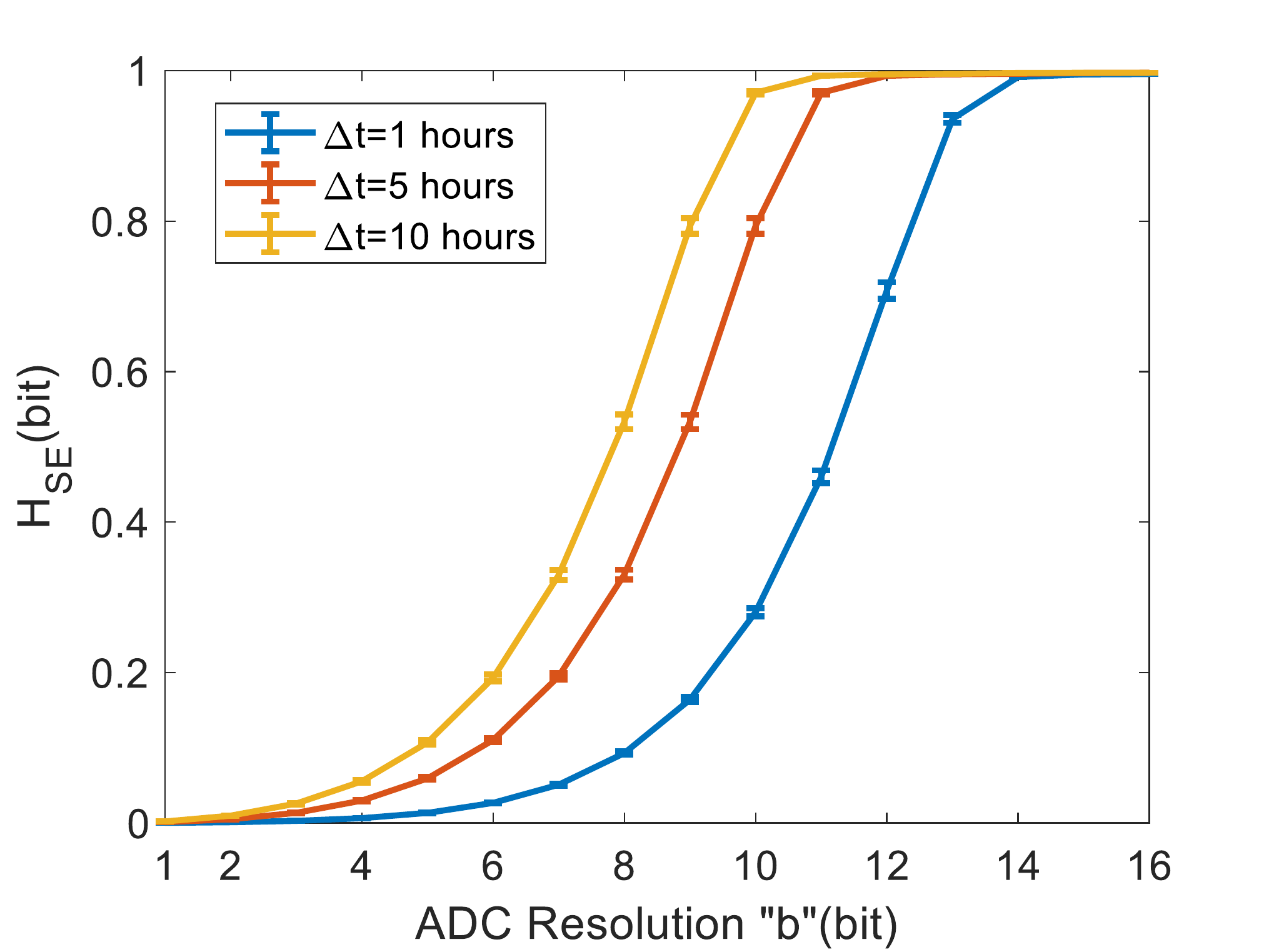}
		\end{center}
		\caption{Uncertainty per bit measured for three different waiting periods $\Delta t$ as a function of the resolution $b$ of the ADC used for measuring the state of the 'key' timers. }
		\label{fig_wait}
	\end{figure}
	\par Next, we consider the information that is available to an attacker and investigate whether he can gain any advantage while predicting the key-string with the information available to him. So, here we note down all the information and resources about the key exchange framework that is potentially available to an attacker:
	\begin{enumerate}
		\item[I1:] We assume that the attacker can passively eavesdrop on the communications over the public channel. This means that the attacker would know which timers were used for a particular key-string.
		\item[I2:] We also assume that the attacker has access to the hardware chipsets.
		\item[I3:] The attacker knows the underlying principle of the timers’ behavior and other logistics of the protocol as described in this work.
		\item[I4:] The attacker has access to a fully functioning quantum computer.
	\end{enumerate}
	\par Now considering I1 and I2, a potential attack could be launched by the adversary where they sample the timers on their copy of the chipset as soon as the user broadcasts a tuple $(\mathbf{O},\mathbf{H},t)$ over the public channel. However, the key that the attacker generates will be at a time instant $t + \Delta t$, where $\Delta t$ is the time that the user waits after they have generated the key. Since the timer values are dynamic in nature, the key generated by the attacker $\mathbf{K_E}$ will be different from the key generated by the user $\mathbf{K_B}$. To quantify the disparity between the keys, we use Shannon information entropy to measure how much information can the attacker gain about $\mathbf{K_B}$ using their own key $\mathbf{K_E}$.
	The average Shannon information entropy contained in each bit generated by the attacker can be expressed as
	\begin{equation}
	H_{SE} = -d\log_2d – (1 – d)\log_2(1 – d)
	\end{equation}
	where $d$ is the average difference in bits between $\mathbf{K_B}$ and $\mathbf{K_E}$. The parameter $H_{SE}$ quantifies the uncertainty of the attacker for every bit of the key $\mathbf{K_B}$ that he or she tries to predict using $\mathbf{K_E}$. When $d = 0$ i.e. the attacker generates the same key as the user, the information entropy of the attacker is zero, this is because the attacker can predict the key with perfect certainty. A similar argument can be made for the other extreme scenario, when $d = 1$, as the attacker can simply invert each bit that he or she generates and produce $\mathbf{K_B}$. The entropy $H_{SE}$ is also equal to 0 in this case. On the other hand, when $d = 0.5$ exactly half of the bits of $\mathbf{K_E}$ do not match with $\mathbf{K_B}$. This means that if the attacker were to randomly guess all the key-bits they would, on average, end up with the same number of matched bits. Therefore, the attacker has 1 bit of uncertainty for every bit generated and zero information gain on the key. The entropy $H_{SE}$ thus takes the maximum value of 1 in this case. 
	\begin{figure}
		\begin{center}
			\includegraphics[page=1,scale=0.45]{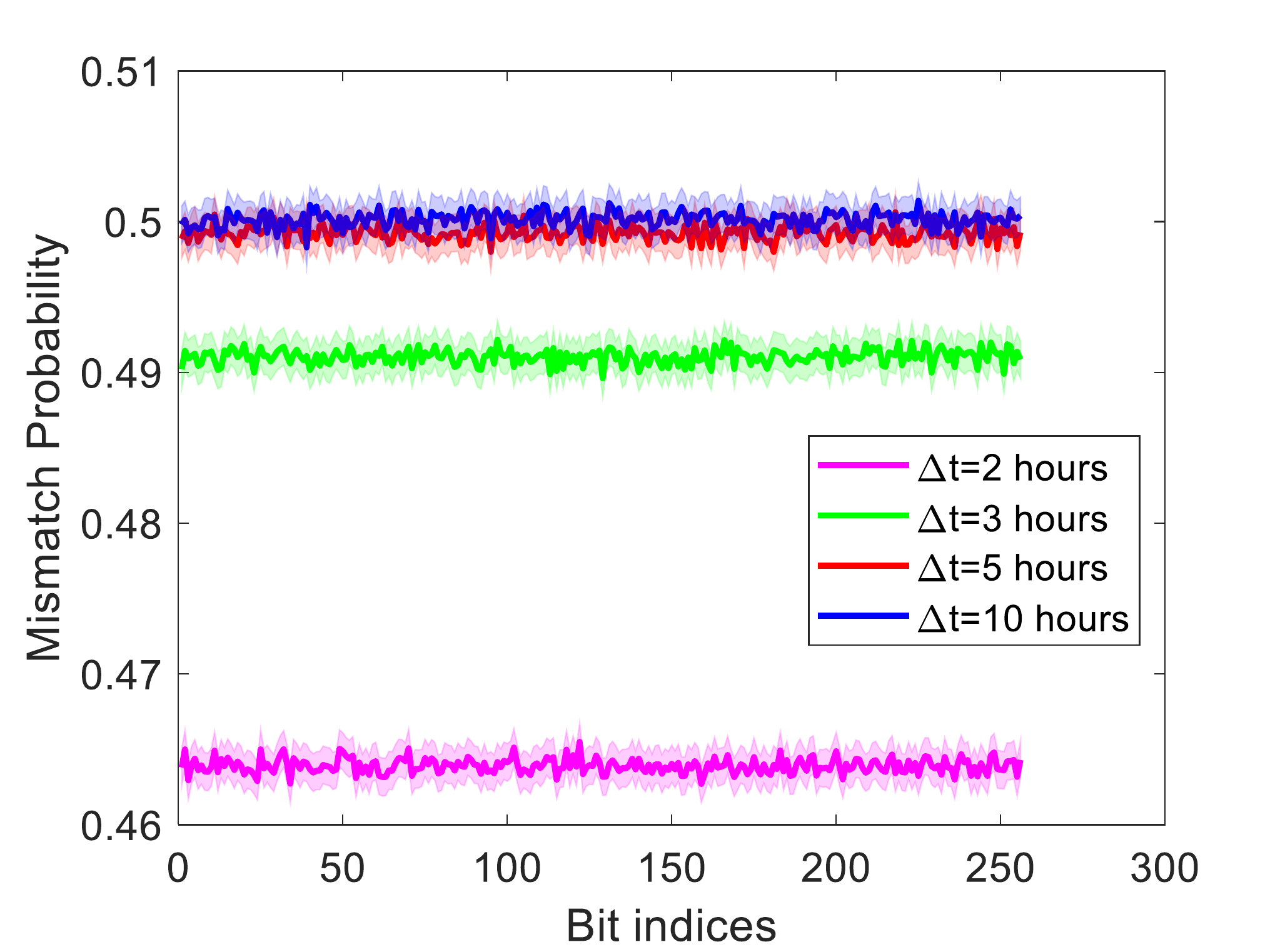}
		\end{center}
		\caption{Probability that the binary states of a timers used in key generation has changed after the waiting period $\Delta t$ hours. Here the resolution of the ADC used for key generation is b=12-bits. The variance across different Monte-carlo trials are highlighted by the shaded region. }
		\label{fig_prob}
	\end{figure}
	\par In order to mimic such a kind of attack we sampled a set of timers and generated keys at random time instances, representing the user’s key, and also sampled the same set of timers at a later instant, which represents the attacker’s key. After that, we calculated the entropy for each sample. Figure~\ref{fig_wait} shows the average uncertainty per bit generated by the attacker when he or she samples the same timer array used by the user. We can observe from the figure that for keys generated with high-resolution ADC, the attacker has almost 1 bit of uncertainty per bit. This means that the attacker is unable to gain any information about the user's key from sampling their own timer chipset.
	The overall trend for the curves with the same wait period (which corresponds to $\Delta t$) can be explained by the fact that at higher resolution, the LSB contains minimum information about the whole dynamic response of the timer. Moreover, the LSB changes much more frequently, and therefore key generated using LSB is more difficult to predict for the same waiting period. It gets increasingly easier to predict as the resolution of the ADC is decreased since the LSB changes slowly and sampling yields more information.
	
	The uncertainty can be increased for a lower-resolution ADC by increasing the waiting period $\Delta t$ which is shown in Figure~\ref{fig_wait} where the curve shifts towards the left as we increase $\Delta t$. This is because as $\Delta t$ is increased, the probability that an ADC bit has changed will also increase, thereby sampling the bits will not provide any useful information. 
	\par 
	However, average Shannon information entropy $H_{SE}$ is agnostic of the position of the mismatched bits.  For instance, one pathological case could be that always the first half of the key-string obtained by the attacker is mismatched while the second half always matches with the true key-string. In this scenario, HSE would still be 1, but the attacker can easily guess the correct key-string. In our next analysis, we show that the probability of such a case is negligible (practically does not exist). Using Monte Carlo simulations with ADC resolution b=12-bits, we counted how many times each of the key-bits in the 256-bit key string gets mismatched among all the iteration and calculated the probability of mismatch for each bit index. Figure 8 shows the probability of mismatch for each bit index after different waiting periods. We can observe that as the waiting period increases each bit index has an approximately equal probability of 0.5 for being mismatched. This shows that there is no bias with respect to the positioning of the mismatched bits and each key bit generated by the attacker has an equal probability of being correct or incorrect which is the same as purely guessing. For a lower waiting period the probability of mismatch decreases for all the bit indices which is in accordance with our previous analysis, but the mismatch probability is approximately the same irrespective of the bit position. Thus, as long as a reasonable resolution ADC is used for measuring the state of the timer and the waiting period is large enough, the attacker will not be able to predict as to what key string was generated by a user. Therefore, I1 and I2 do not reveal any information about the secret key and the attacker would still need to resort to brute force search for a successful attack.
	
	So far we have shown that the key exchange protocol is secure based on the facts that the keys used are completely random in nature and from the public information available during the key exchange the attacker can not gain any information about the random keys. Next, we consider I3 available to an attacker and investigate whether they could predict the keys by using their knowledge about the timer behavioral (or software) model. However, since they do not have access to the timer initialization parameters $\overline{P_i}$, they cannot use the public information $(\mathbf{O},\mathbf{H},t)$ to decipher the states $s_{O_l}(t)$. Also, the attacker is unable to rewind the hardware timer on their copy of the chipset to measure the states $s_{O_l}(t)$ going back in time. Therefore, the only way to predict $\mathbf{K_B}$ would be to solve equation~\ref{eqn:keystring} for each bit of the key for finding the secret parameters $\overline{P_1}, \overline{P_2}...\overline{P_N}$. In the next set of analysis, we will show that it is practically impossible to find the secret parameters from the hardware chipsets themselves.
	\par First, we consider equation 1 where the parameters could be regressed if a timer is sampled multiple times to measure $I_{timer}(t)$ at different time instances. However, this is only true if the attacker can get access to the precise value of $I_{timer}(t)$. From equation 2 we observe that the binary state of the timer only provides a single bit of information about $I_{timer}(t)$. Moreover, axiom SP4 dictates that even the single bit of information about $I_{timer}(t)$ is XOR-ed with other $G$ hash timers' binary states. The attacker has only access to the XOR-ed output due to the manner in which hardware chipsets are designed. Therefore each bit of key-string the attacker samples from the hardware chipsets will be derived from $G+1$ timers. Note that there is no analytical solution for equation~\ref{eqn:keystring} so the attacker will have to resort to a brute-force numerical search. We now show how the SPoTKD protocol is secure against such attacks under the standard model.
	\par\textbf{Claim 1.} The SPoTKD protocol is secure under the standard model.
	\begin{proof}	
	\par$\>$ Each key bit $Q_L(t)$ is derived from the temporal responses of $G+1$ timers where $G$ is the number of hash timers used in key generation. Now, the temporal response of each timer is determined by the secret parameter tuples $\overline{P} = [p_0, p_1, p_2, p_3]$. Therefore, each key bit, in turn, is determined by $G+1$ tuples of $\overline{P}$. We define $p_{Total}$ as the total number of parameters from which each bit is derived which is given by
	\begin{equation}
	p_{Total} = 4(G+1)
	\end{equation}
	This means that the search space would be a matrix with $p_{Total}$ dimensions. Now, let $R$ be the range of possible values for each of the $p_{Total}$ parameters. Then the total number of elements in the matrix i.e. the total search space $SP_{Total}$ would be given by
	\begin{equation}
	SP_{Total}=R^{4(G+1)}
	\label{eqn:space}
	\end{equation}
	Even though the parameters $\overline{P}$ are determined by the timer initialization conditions and timer form factors, they are calibration parameters. Assuming a double-precision floating-point for the parameters implies that $R = 2^{63}$. This yields
	\begin{equation}
	SP_{Total}=2^{252(G+1)}
	\label{eqn:realspace}
	\end{equation}
	For $G = 128$ hash timers (which was used in our simulations for generating the key string) this would result to a search space of $2^{32508}$ possible combinations. Therefore, an attacker employing a brute-force search strategy would require $2^{32508}$ bits of storage, which is prohibitively large. Moreover, even if the attacker uses the fastest computer in the world \cite{fugaku2020}, which can perform $10^{19}$ computations per second, it will take them approximately $2^{32444}$ seconds, or $2^{32419}$ years to search the entire space. Since we assumed that the attacker is only constrained by the computational/storage resources and time available to them, hence, under the standard model, the SPoTKD protocol is secure.
	\end{proof}
	\par Next, we consider I4 where we assume that the attacker has access to a quantum computer with large enough storage space and computational resources to search the aforementioned solution space in a reasonable amount of time. In this analysis, we show that our protocol remains secure if we impose a physical constraint that limits the number of hardware chips that the attacker can use for measurement.
	\par\textbf{Claim 2.} The SPoTKD protocol is resistant to quantum attacks.
	\begin{proof}

	\par$\>$ Equation~(\ref{eqn:keystring}) has no unique solution and since the parameters are randomly chosen by the server, every solution within the search space is equally likely to be the correct one. The only way to eliminate possible combinations from the solution set would be to sample each hardware timer at multiple time instances and solve equation~\ref{eqn:keystring} repeatedly. Since equation~\ref{eqn:bitgen} is symmetric the expected size of the solution set, denoted as $\mathbb{E}(SP_{J})$, after each sampling reduces by 
	\begin{equation}
	\begin{aligned}
	\mathbb{E}(SP_{J})&=\frac{SP_{Total}}{2^J}\\
	&=\frac{2^{252(G+1)}}{2^J}
	\end{aligned}
	\label{eqn:sampspace}
	\end{equation}
	where $J \in \mathbb{Z}^+$ indicates the number of samples. This means that if the attacker can sample each timer enough number of times, they can find out the initialization parameter $\overline{P}$. However since the timers are designed for one-time read (Axiom SP5 in section \ref{back}), the attacker is unable to make multiple measurements on a timer using the same chipset. For each measurement, the attacker would therefore require a new chipset. Thus, there is an upper bound to the number of measurements that an attacker can perform, which is the total number of chipsets $C_{Total}$ available. Therefore we have
	\begin{equation}
	J\leq C_{Total}
	\label{eqn:samplebound}
	\end{equation}
	 Now if we constrain the total number of chipsets $C_{Total}$ according to
	\begin{equation}
	C_{Total}<252(G+1)
	\label{eqn:chiprelation}
	\end{equation}
	then the attacker would still be unable to find the unique solution to equation~\ref{eqn:keystring} since 
	\begin{equation}
	\hspace{15mm}\mathbb{E}(SP_{J})>2 \hspace{10mm} \forall J 
	\label{eqn:expectedsize}
	\end{equation}
	 Note that, the constraint here for an attacker is not the computational power available to them but rather the physical resources they can acquire. Thus, the key exchange protocol is resistant to quantum attacks.
	\end{proof}
	\begin{table}
		\begin{center}
			\caption{Performance Comparison between SPoTKD and other state-of-the-art key exchange protocol}
			\label{tab:table1}
			\begin{tabular}{c|c|c|c|c}
				& \textbf{Key}& \textbf{Security} & \textbf{Computational} & \\ 
				\textbf{Protocol}& \textbf{Length}& \textbf{Strength}&\textbf{Cost}&\textbf{Scalability}\\
				& (bits)& (bits)& (no. of cycles)&\\
				\hline
				&&&&\\
				PPUF\cite{PPUF} & 1024 & 112 & $\mathcal{O}(10^{16})$ & High\\ 
				\hline
				&&&&\\
				QKD\cite{Ben84} & 256 & 256 & $\mathcal{O}(10^{4})$ & Low\\ 
				\hline
				&&&&\\
				RSA\cite{Riv78} & 3072 & 128 & $\mathcal{O}(10^{7})$ & High\\ 
				\hline
				&&&&\\
				\textbf{SpoTKD} & \textbf{256} & \textbf{256} & $\mathbf{\mathcal{O}(10^{2})}$ &\textbf{High}\\ 
			\end{tabular}
		\end{center}
	\end{table}
	In the next set of analysis we want to show how the proposed key exchange protocol is secure against most popular kind of attacks.
	\par\textbf{Claim 3.} The proposed protocol is secure against man-in-the-middle attacks.
	\begin{proof}
		During the SPoTKD protocol, a user publicly broadcasts the tuples $(\mathbf{O},\mathbf{H},t)$ indicating the timer indexes the user sampled along with the time at which they were sampled. For an attacker to successfully impersonate the server, they will need to know the secret timer parameters $\overline{P}$, which is never revealed during any phase of the protocol. Also, our previous analysis shows that it is practically impossible to find out these parameters using brute-force search. Note that all the publicly distributed chipsets store the same information on the timers and authentication is carried out only after the server and user have established a secure channel subsequent to a successful key exchange. Thus, the attacker cannot impersonate any user.
	\end{proof}
	\par\textbf{Claim 4.} The proposed protocol is secure against replay attacks.
	\begin{proof}
	Once a set of timers is used for key exchange, they are desynchronized with respect to the server’s model (Axiom SP5 in section~\ref{back}). Thus, during every session, a new set of timers is used to exchange keys. This means that a new key is generated for every new session. Also, during the key exchange protocol, the measured states of the timers are never made public. Therefore, the attacker cannot use any information from previous sessions to their advantage. This implies that the SPoTKD protocol is secure against replay attacks.
	\end{proof}
	\par\textbf{Claim 5.} SPoTKD protocol is secure against backward and forward traceability attacks.
	\begin{proof}
	In our protocol, the keys generated are random in nature as shown in figure~\ref{fig_random} that are not predictable. Also, each key is used only once. Therefore the key exchange at session instance $SS_a$ can not be inferred from other keys at any other session $SS_b$, where $a\neq b$. Moreover, we have shown in the previous claims that inferring any knowledge about the secret parameters is also practically impossible. Therefore, the SPoTKD protocol is immune to forward or backward traceability attacks.
	\end{proof}
	
\begin{figure}
	\begin{center}
		\includegraphics[page=1,scale=0.45]{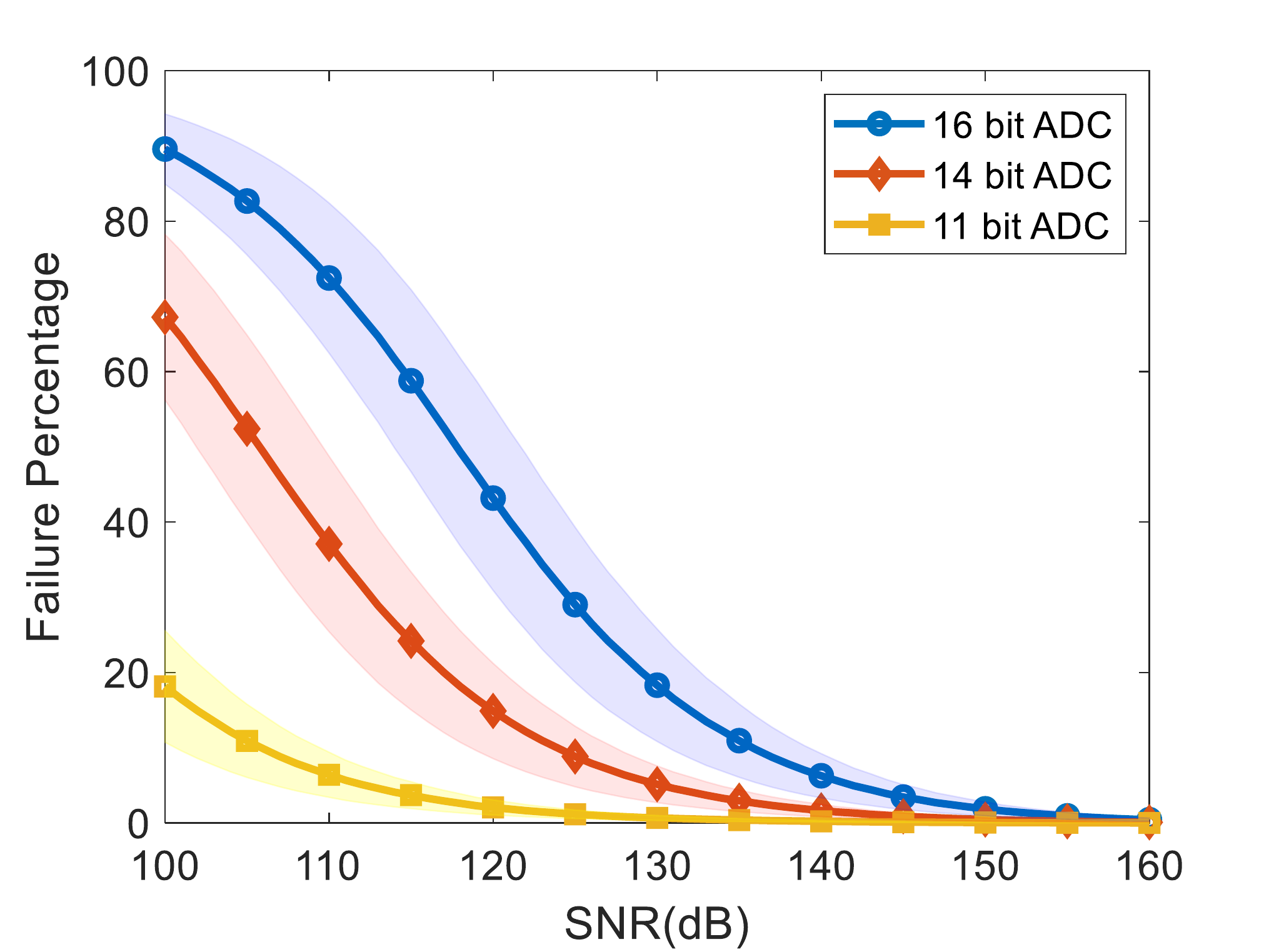}
	\end{center}
	\caption{Improvement in noise-robustness of the SPoTKD protocol when the resolution $b$ of the ADC used for measuring the state of the 'key' timers is decreased. The variance across different Monte-carlo trials are highlighted by the shaded region.  }
	\label{fig_highbit}
\end{figure}
	\par\textbf{Claim 6.} SPoTKD protocol is resistant to de-synchronization attacks.
	\begin{proof}
		The robustness of the timer response ensures that the dynamics of the hardware timer remain synchronized with its software model on the server. According to Axioms SP1-SP4 in section~\ref{sec:primitives}, the timer’s dynamic response on any user’s chip cannot be programmed or altered by the attacker unless and until the attacker gets access to the chip physically. In such a case where the user suspects that his or her chip may have been compromised physically by an attacker, the user can simply discard the chip and procure a new one, since all the chipsets have the same information that is stored.  Thus, the protocol is resistant to de-synchronization attacks.
	\end{proof}
	In addition, the construction, operating principle and inherent security of the quantum-tunneling device i.e. the self-powered timer \cite{Zho17a} also prevent the attacker to probe the state of the timer by using any side-channel (power or electromagnetic) without affecting the state of the timer (Axioms SP1-SP6 in section~\ref{sec:primitives}). Therefore, in this regard, the timer chipset emulates a quantum communication channel \cite{Ngu07}, but using an analog dynamical system that is secure against any side-channel attacks.
\begin{figure*}[b]
	\begin{center}
		\includegraphics[page=1,scale=0.52]{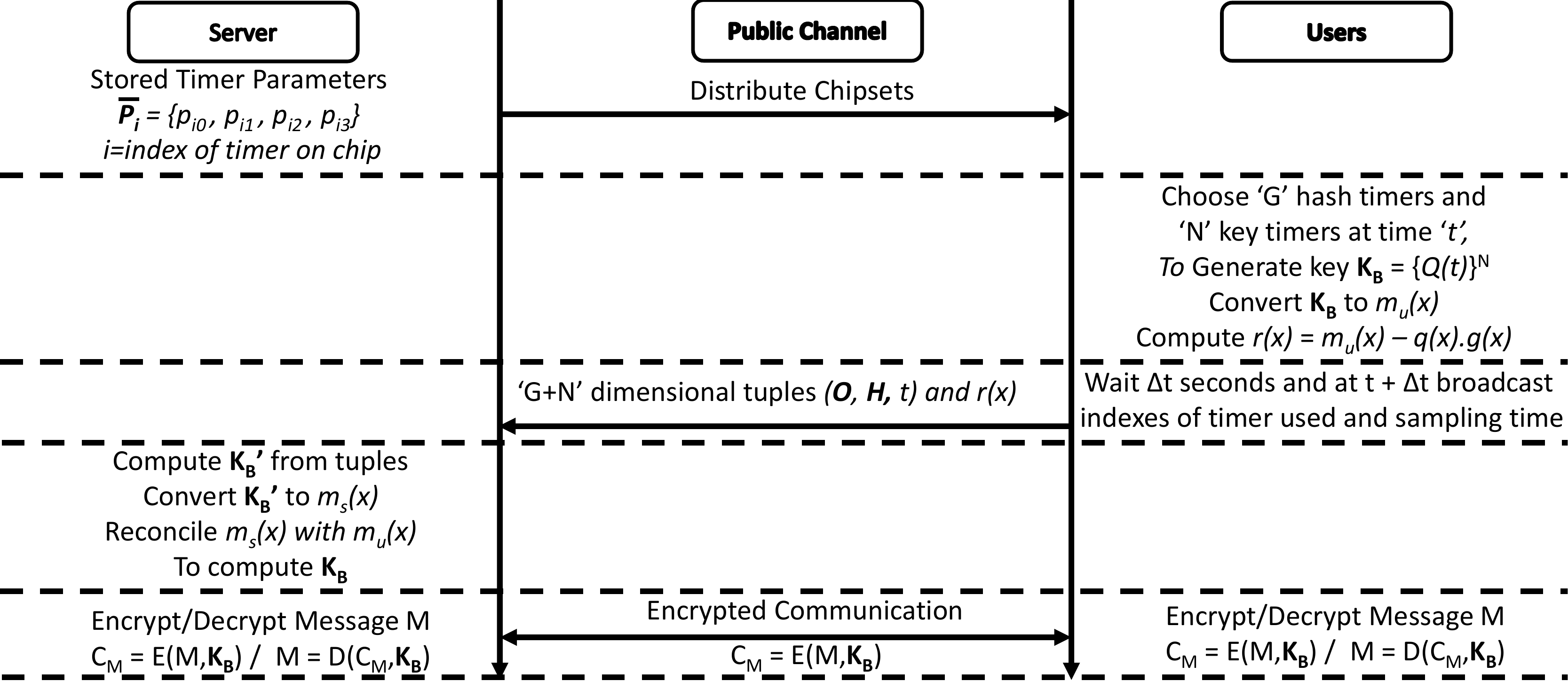}
	\end{center}
	\caption{Modified SPoTKD protocol between a server and a user incorporating error-correction }
	\label{protocol_3}
\end{figure*}	
	\par We have evaluated the performance of our proposed protocol with similar hardware-software based key exchange protocols such as PPUF~\cite{PPUF} and some state-of-the-art key exchange protocols such as RSA~\cite{Riv78} that are currently being used. The comparison is summarized in Table 1 with respect to criteria such as key length, security strength, computational cost, and scalability. Here security strength measures the number of trials required to brute-force a key irrespective of the key length. A $128$-bit security means $2^{128}$ trials to break the protocol. We also compared the computational resources required to perform a single key exchange in terms of the number of computation cycles. And finally, scalability indicates the ease at which the key exchange protocol can accommodate large of number of users. Since our goal is to provide secure key exchange among a large number of users using low resources, these features are extremely important to evaluate and compare different designs.
	\par We start by evaluating the security strength of each protocol. For PPUF using 1024 bit key, an attacker needs to perform $1.7x10^{29}$ cycles of simulation on average to find the secret key~\cite{PPUF}. Accounting for overhead computation this roughly translates to a 112-bit security. According to NIST 2020 recommendations, RSA requires a key length of 3072-bits to achieve a security strength of 128-bit. Now, since both QKD and SPoTKD use symmetric key encryption (AES), a 256-bit key length corresponds to a security strength of 256-bit. Due to the use of a large key size, both PPUF and RSA are computationally expensive. The PPUF based key exchange protocol requires approximately $10^{16}$ cycles of computation~\cite{PPUF} and RSA requires $\mathcal{O}(10^7)$ computational cycles~\cite{RSAcomp}. Even though QKD uses a much smaller key-string, additional computation needs to be performed for the error reconciliation protocol. The computational complexity is of order $\mathcal{O}(10^4)$ for a 256-bit key using common error-correcting code~\cite{qkdcomp}. Meanwhile, for the basic SPoTKD the user needs to simply measure the state of the timers once and perform $G=128$ bit-wise XOR from the hash timers to generate the bit $X(t)$. This can be done in $log_2(G)$ computational cycles. After that, the outputs of the $N$ key timers are XOR-ed with the bit $X(t)$ in $N$ cycles. In this regard, our protocol is by far the most efficient. If error correcting SPoTKD is used(discussed in Section VII), the computational cost will be similar to that of QKD. In addition, we have shown in our analysis that our protocol is resistant to quantum attacks, similar to QKD. In comparison, however, QKD is expensive and in its current state is not portable or scalable to support a large number of users. On the other hand, our protocol is based on silicon fabrication technology which is relatively inexpensive at a production scale and the fabricated chipsets can be easily distributed to millions of users.


	\section{Noise Robustness}
	\label{noise}

	In the next set of experiments, we quantified the robustness of the SPoTKD protocols in the presence of real-world operational artifacts. For instance, the timer on a physical chip could inadvertently desynchronize with the software model on the server.  This could be due to fabrication mismatch, environmental variations, device degradation, and measurement noise. To emulate this effect we performed a Monte Carlo study where we added White Gaussian Noise to the timer response and then generated the keys by sampling at random time instances. 
	In this case, the SNR is defined as
	\begin{equation*}
	SNR=\frac{P_{Signal}}{P_{Noise}}
	\end{equation*}
	where $P_{Signal}$ is square of the signal output measured from the timer and $P_{Noise}$ is the signal variance. This ‘measured’ key was compared against the ‘gold’ key generated from the software model in the server i.e. without any noise. Every instance where the keys do not match perfectly is counted as a failure. Figure~\ref{fig_highbit} shows the failure percentage, calculated as the average number of failure instances over all the instances of simulation, at each noise level. As expected, the failure percentage reduces with an increase in SNR. 
	\par Better noise robustness could be achieved by using low-resolution ADC for the key timers, as shown in Figure~\ref{fig_highbit}. However, as we have shown in the previous section this could lead to more information gained by a ‘knowledgeable’ attacker to predict the key. In order to mitigate this threat, the server can recommend the user to opt for an increase in the wait-period $\Delta t$ and achieve the same level of uncertainty even for low-resolution ADC, as illustrated in Figure~\ref{fig_wait}. Thereby, a tradeoff exists between the level of security and the waiting period, and the preference for one or the other depends on the target application. 
	
	\section{Error Correcting SPoTKD}
	\label{errcor}
	\begin{figure}
		\begin{center}
			\includegraphics[page=1,scale=0.45]{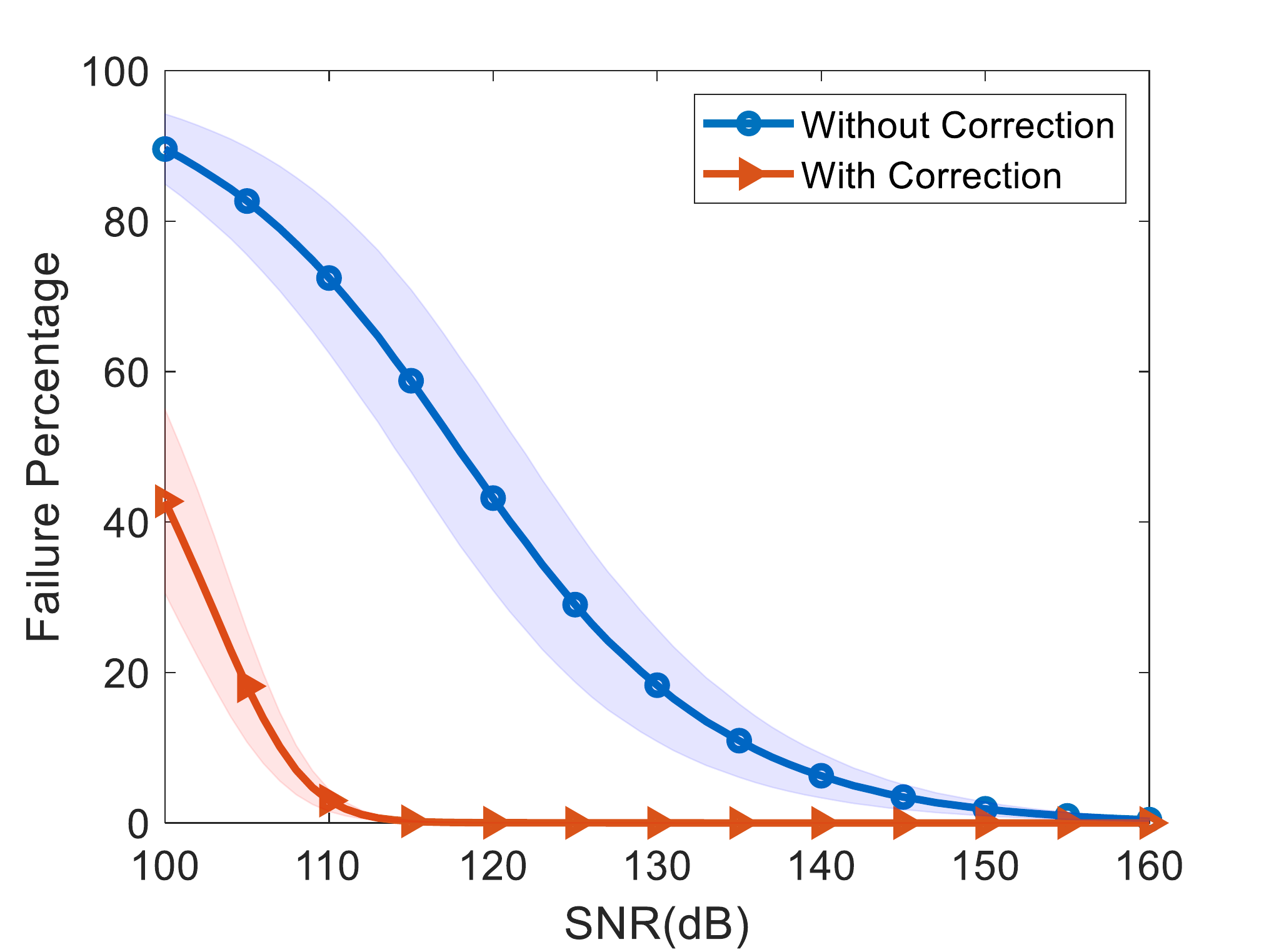}
		\end{center}
		\caption{ Performance of the SPoKTD protocol in the presence of noise when error-correction is used. A 16-bit ADC was used to measure the state of the 'key' timers. The variance across different Monte-carlo trials are highlighted by the shaded region. }
		\label{fig_cor}
	\end{figure}
	In the previous section, we have discussed how the protocol’s robustness to noise could be increased by either trading off security or waiting period. In this section, we will discuss a new protocol shown in Figure~\ref{protocol_3} in which noise robustness can be improved without compromising neither security nor waiting time by using standard error-correcting codes which are generally used in digital communication. For our purpose, we will use cyclic-redundancy-check (CRC) for error correction~\cite{crc}, even though other error-correcting codes could also be used. 
	
	\par The string of key-bits are represented as the coefficients of a message polynomial, $m(x)$, over a Galois field (GF2) and to find the CRC, the message polynomial is multiplied by $x^n$  and then the remainder $r(x)$ is found by dividing with an n-degree generator polynomial $g(x)$. The coefficients of the remainder polynomial are the bits of the CRC. This can be expressed as
	\begin{equation}
	m_u(x).x^n=q\small(x\small).g\small(x\small)+r\small(x\small)
	\end{equation}
	
	where $q(x)$ is the quotient. Typically, $m_u(x).x^n – r(x)$ and $g(x)$ is sent over the communication channel. However, in this protocol we are sending $r(x)$ i.e. only the CRC bits together with the tuples $(\mathbf{O},\mathbf{H},t))$ over an insecure channel as illustrated in Figure~\ref{protocol_3}, and $g(x)$ is assumed to be predetermined and a public knowledge. This is because we do not want to share the message $m_u(x)$ which is the key itself. The server generates the $m_s(x)$ using the tuples $(\mathbf{O},\mathbf{H},t))$ information and the software model. Then, together with $r(x)$ and $g(x)$ the sever can reconcile $m_s(x)$ with $m_u(x)$ up to a certain hamming distance. Thereby, tolerating erroneous key-bits measured by the user due to noise.
	\begin{figure}
		\begin{center}
			\includegraphics[page=1,scale=0.45]{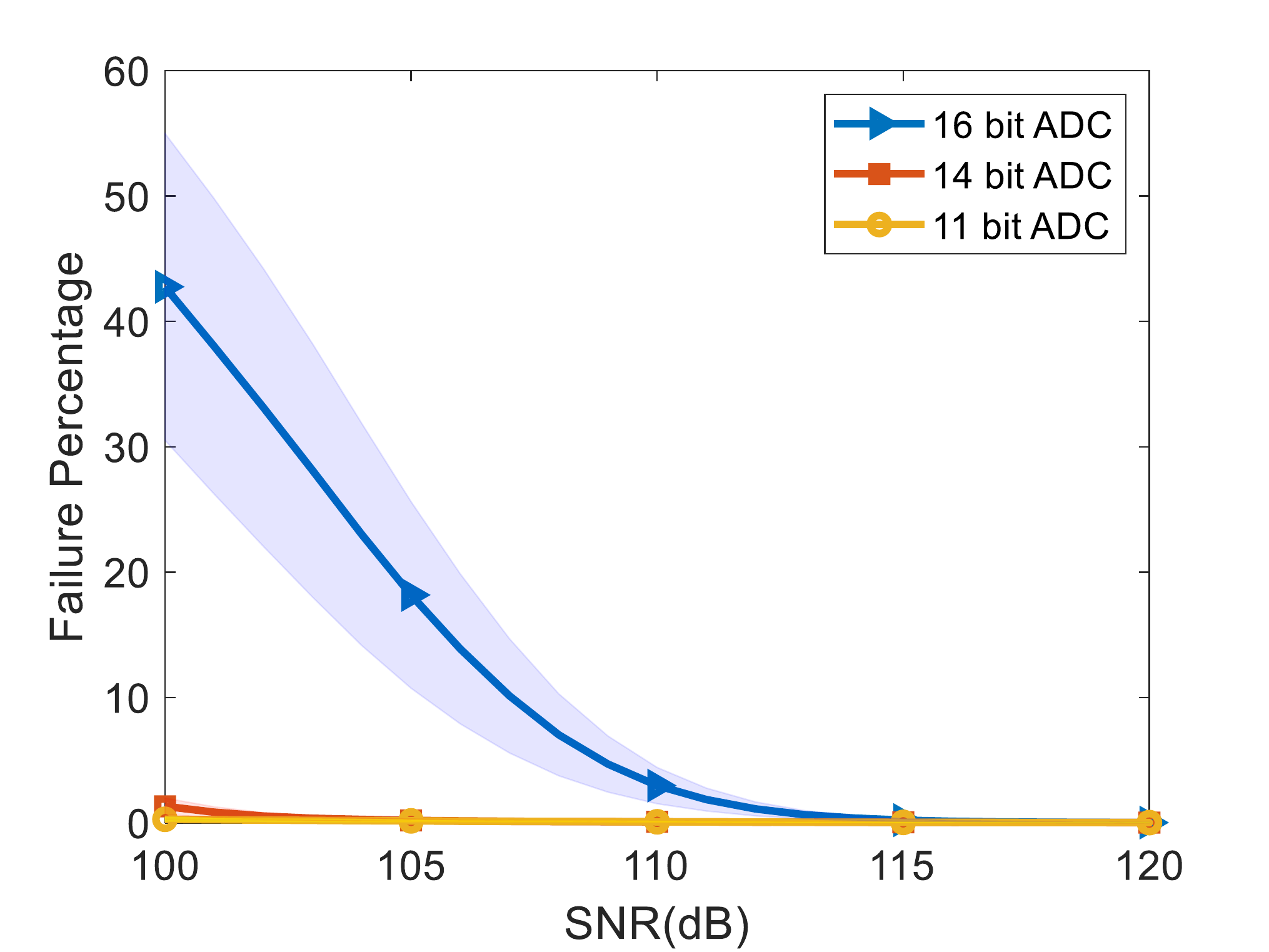}
		\end{center}
		\caption{ Performance of the SPoTKD protocol in the presence of noise when using error-correction and when the resolution $b$ of the ADC used for measuring the state of the 'key' timer is reduced. The variance across different Monte-carlo trials are highlighted by the shaded region. }
		\label{fig_highcor}
	\end{figure}

	\par From the security point of view, the attacker now has more information about the key as the remainder $r(x)$ is broadcast along with the $(\mathbf{O},\mathbf{H},t))$ tuples. For example, let $m(x)$ be the representation of a 256-bit key. Then the number of possible keys = $2^{256}$. We assume that the attacker has an identical chip himself. Let $g(x)$ be a 28-degree polynomial, then with the knowledge of $r(x)$ the number of possible keys is reduced to $2^{256-28} = 2^{228}$. Therefore, the search complexity for an attacker decreases proportionally to the degree of generator polynomial used i.e. number of CRC bits.
	\par In order to counteract this effect, the length of the key can be increased by an amount equal to the degree of $g(x)$. This would mean more timers are needed to be used for an effective key length equal to the number of timers used minus the degree of $g(x)$. In the example described above, the number of timers required for a 256-bit effective key length would be 284. 
	
	According to Philip Koopman’s table of CRC generator polynomial~\cite{table}, for a $g(x)$ of $28$ degrees and data-word length less than $483$ bit, the least hamming distance that can be corrected is $8$. Therefore, we can allow up to $8$ mismatches for the 284-bit key, which has an effective key length of 256-bits, and then compare the noise robustness to the 256-bit key. This is illustrated in Figure~\ref{fig_cor} which shows significant noise robustness improvement. This is achieved without sacrificing any complexity and does not come at the cost of a longer waiting period. Robustness can be further improved by using lower resolution ADC for key-generation as shown in Figure~\ref{fig_highcor} if the user opts for more accuracy and is compliant with a longer waiting period.
	
	\section{Discussions and Conclusions}
	\label{disc}

	\par In this paper, we introduced a novel key distribution framework, SPotKD, based on specific security features of the previously reported self-powered time-keeping devices. We described the key exchange protocol and also analyzed it both from a security and noise robustness point of view. Our protocol is not only secure against most kinds of attacks but also proved to be secure in the advent of a fully functional quantum computer in the future. We have also evaluated the performance of our protocol against some state-of-the-art key distribution schemes.
	\par Several challenges exist in implementing the proposed key distribution system from a practical point of view. At the core of the system is the self-powered timer technology which have been successfully demonstrated in our prior work \cite{Zho17a,Zho19}. However, designing the peripheral circuitry that can realize the key generation protocol on-chip is yet to be accomplished. A complete system-on-chip (SoC) should consist of an array of these timers and a combinational logic circuit that will allow the user to arbitrarily choose any set of timers for key generation. In addition, the circuit for destroying the timer’s information should also be integrated into  the chipsets. Furthermore, the design of the destruction circuit should be done in such a manner that the timers’ temporal response becomes desynchronized even before the user can access the output of the chipsets and remain desynchronized for a significant period after read-out. Only then, the timers can be considered to be a one-time read device. Addressing this challenge would be a part of future research.
	\par Another limitation arises in scaling the framework due to the limited number of chips that can be distributed while maintaining security against quantum attacks as discussed in claim 2. However, the limit on the number of chipsets can be increased by using more hash timers during key generation. It should also be noted that due to real-world artifact noise, increasing the number of hash timers may lead to high failure rates during key exchange. The protocol will remain secure albeit slight increase in the probability of a key exchange failure and a trade-off exists  between the security and the reliability of the SPoTKD protocol. In this regard, incorporating  error correcting mechanisms in the SPoTKD protocol will help to address these limitations.
	
	\par One other limitation to consider is that the underlying assumption for SPoTKD dictates that the server has ample resources to securely store the timer parameters and to secure access control. With respect to secure storage, the server can adopt traditional, high-end and computationally intensive symmetric key encryption approaches. However, the protocol in its current state will not remain secure if the server becomes compromised and attacker gain access to the timer initialization parameters using phishing techniques or by compromising the access control protocols (similar to the attack models demonstrated for trusted program modules~\cite{tpm2021}). This vulnerability can be overcome by adopting a distributed server (Decentralized Cloud Storage) approach. The security of these types of storage system is well established~\cite{decent} where AES-256 is used to encrypt the data and then each data is split and stored across a distributed network. Another solution that we are currently investigating, is storing the timer initialization parameters in a semi-persistent storage (memory whose content is destroyed after a pre-determined time). This attribute will prevent against the “record now decode later” attacks where the attacker logs the encrypted data with the hope that a powerful computer will be available to successfully decrypt the data, or the server storage will be compromised at a much later time.
	
	\par Our future work would focus on prototyping a self-powered timer system-on-chip with all the basic hardware security primitives. We will then validate the SPoTKD protocol under real-world conditions and over different distribution channels. This will open the possibility of applying SPoTKD in areas such as quantum secure blockchains (based on symmetric-key) and electronic voting.
	\section*{Acknowledgments}
	The authors would like to thank Dr. Kenji Aono, Dr. Darshit Mehta and Dr. Sri Harsha Kondapalli at the Electrical and Systems Engineering department, Washington University in St. Louis, for their valuable assistance in running experiments and MATLAB$^\copyright$ simulations.
	
	\appendix
	\section{Derivation of Equation 1}
	
	In \cite{Zho17a} it was shown that the floating gate(FG) potential of the timer can be described using a first-order differential equation   
	\begin{equation}
	V_{FG} \small(t\small)=\frac{\beta t_{ox0}}{ln\small(p_1t + p_0\small)}
	\label{eq_resonant}
	\end{equation}
	where $V_{FG}(t)$ is the potential of the timer at time instant $t$ and $p_0$ and $p_1$ are the model parameters defined as
	\begin{equation*}
	p_0=\exp\Big(\frac{\beta t_{ox0}}{V_0}\Big),p_1=\frac{A_0\alpha\beta}{C_T t_{ox0}}
	\end{equation*}
	Here $V_0$ refers to the initial voltage, $A_0$ is the tunneling junction area, $t_{ox0}$ is the average oxide thickness of the device and $C_T$ is the total capacitance of FG. $\alpha$ and $\beta$ are functions of the material properties and depends on the technology used for fabricating the timers. When the floating gate is coupled to the gate of a readout MOSFET which is biased in weak-inversion, then its drain to source current($I_{timer}(t)$) can be expressed as
	\begin{equation}
	I_{timer}\small(t\small)=I_0\exp\Big[\frac{K_p\small(V_s-V_{FG}(t)-V_T\small)}{U_T}\Big]
	\end{equation}
	where $V_s$ is the source voltage, $V_T$ is the threshold voltage of the MOSFET, $I_0$ represents a characteristic current of the MOSFET, $K_p$ is the gate efficiency and $U_T$ is the thermal voltage. Substituting $V_{FG}(t)$ from equation~\ref{eq_resonant} we get
	\begin{equation}
	I_{timer}\small(t\small)=p_3\exp\Big[-\frac{p_2}{ln\small(p_1t + p_0\small)}\Big]
	\end{equation}
	where $p_2=\frac{K_p}{U_T}\beta t_{ox0}$, and $p_3=I_0\exp\Big[\frac{K_p\small(V_s-V_T\small)}{U_T}\Big]$
	\bibliographystyle{ieeetr}
	\bibliography{references}

	
\end{document}